\documentclass[aps,prl,reprint,floatfix,amsmath,amssymb]{revtex4-2}

\usepackage{epsfig}
\usepackage{graphicx}
\usepackage{dcolumn}
\usepackage{bm}
\usepackage[colorlinks=true,citecolor=blue, dvipdfm]{hyperref}
\usepackage{amssymb}

\begin{document}
\title{Is there Kibble-Zurek scaling of topological defects in first-order phase transitions?}

\author{Fan Zhong}
\affiliation{School of Physics and State Key Laboratory of Optoelectronic Materials and
Technologies, Sun Yat-sen University, Guangzhou 510275, People's Republic of China}

\date{\today}

\begin{abstract}
Kibble-Zurek scaling is the scaling of the density of the topological defects formed via the Kibble-Zurek mechanism with respect to the rate at which a system is cooled across a continuous phase transition. Recently, the density of the topological defects formed via the Kibble-Zurek mechanism was computed for a system cooled through a first-order phase transition instead of the usual continuous transitions. Here we address the problem of whether such defects generated across a first-order phase transition exhibit Kibble-Zurek scaling similar to the case in continuous phase transitions. We show that any possible Kibble-Zurek scaling for the topological defects can only be a very rough approximation due to an intrinsic field for the scaling. However, complete universal scaling for other properties does exist. 
\end{abstract}

\maketitle

{\it Introduction.}---The Kibble-Zurek (KZ) mechanism for topological defects formed when a system is cooled through a continuous phase transition into a symmetry-broken ordered state~\cite{Kibble1,Kibble2,Zurek1,Zurek2,revqkz1,revqkz2,Zurek4}, has attracted great attention for many years. It was first proposed in cosmology~\cite{Kibble1,Kibble2} and later applied to condensed-matter physics~\cite{Zurek1,Zurek2}. Upon combining the equilibrium scaling near the critical point with the adiabatic--impulse--adiabatic approximation, a universal KZ scaling for the defect density has been proposed~\cite{Zurek1,Zurek2,Zurek4,revqkz1,revqkz2}. It has since been applied intensively to many systems, ranging from classical~\cite{Laguna1,Laguna2,Laguna3,Laguna4,Laguna5,Laguna6,Laguna7,Laguna8,Laguna9,Laguna10,Laguna11,Laguna12,Laguna13,Laguna14,Laguna15,Vinas1,Vinas2,Vinas3,Vinas4,Vinas5,Vinas6} to quantum phase transitions~\cite{qkz1,qkz2,qkz3,qkz4,qkz5,qkz6,qkz7,qkz8,qkz9,Clark,Weiss,Keesling,Weinberg,Puebla,Schmitt,King,Sadhasivam}.  Recently, the density of topological defects formed via the KZ mechanism was computed for a system cooled through a first-order phase transition (FOPT) instead of the usual continuous transitions~\cite{Suzuki}. However, no scaling of the defect density was considered there. As FOPTs exist far more widespread and are known far more earlier than continuous phase transitions, it is of great interest to know whether there exists KZ scaling of topological defects in this case.

Recently, complete universal scaling in FOPTs was demonstrated~\cite{zhong24,zhong24f}. This is to fully consider the relationship between a model that gives rise to field-driven FOPTs and its derived effective cubic theory that controls scaling behavior. Accordingly, the rescaled curves of an order parameter completely collapse onto each other with universal exponents over a large range of control parameters in both zero- and two-dimensional systems. As temperature-driven FOPTs behave differently from field-driven ones~\cite{liang,liange}, we will adapt the theory to the former case and demonstrate similar complete universal scaling in cooling FOPTs. However, the theory developed indicates that the usual KZ scaling of topological defects in FOPTs can at most be observed with a rough nonuniversal effective exponent. In the following, we will first consider a model of cooling FOPTs and then develop the theory for it, confront it with numerical results and conclude with conclusions.

{\it Model.}---Consider a conventional Ginzburg-Landau functional
\begin{equation}
F(\phi)=\int d^dx\left\{f_6(\phi)+\frac{1}{2}\left(\nabla\phi\right)^2\right\}\label{gl}
\end{equation}
in $d$-dimensional space with the Devonshire mean-field (MF) free energy~\cite{Devon,Devon1}
\begin{equation}
f_6(\phi)=\frac{1}{2}a_2\phi^2+\frac{1}{4}a_4\phi^4+\frac{1}{6}a_6\phi^6-H\phi\label{f4}
\end{equation}
for an order parameter $\phi$ and its ordering field $H$, where $a_2$ is a reduced temperature, $a_4<0$ and $a_6>0$ are coupling constants. Dynamics can be studied by the usual Langevin equation~\cite{Hohenberg} $\partial\phi/\partial t=-\delta F/\delta\phi+\zeta$, or
\begin{eqnarray}
\partial\phi/\partial t=-\left(a_2\phi+a_4\phi^3+a_6\phi^5-H-\nabla^2\phi\right)+\zeta,\nonumber\\
\langle\zeta({\bf x},t)\zeta({\bf x}',t')\rangle=2 T\delta(t-t')\delta({\bf x}-{\bf x}'),\qquad\label{lang2d}
\end{eqnarray}
where $\zeta$ is a Gaussian white noise with a noise strength or temperature $T$ and we have chosen the time unit to fix the usual kinetic constant at 1. In the absence of the ordering field $H$, $f_6$ possesses a temperature-driven FOPT for $a_4<0$ at an equilibrium point $a_2=3a_4^2/16a_6$ at which two degenerate ordered phases with $M_{\rm eq}^2=(-a_4+\sqrt{a_4^2-4a_2a_6})/2a_6$ and a disordered phase with $M=0$ share identical free energy, where $M=\langle\phi\rangle=\phi$ as no ensemble average over the noise is needed for $T=0$. The disordered phase loses its stability at $a_{2s}^{\rm MF}=0$, the MF spinodal, below which spinodal decomposition to the ordered phases ensues~\cite{Gunton83,Binder,Binder2}. Upon taking into account fluctuations, the cooling transition can proceed above $a_{2s}^{\rm MF}$. This results in domains of the two coexistent ordered phases with the domain walls being topological defects. For a continuous transition, the dependence of the density of these defects on the cooling rate is the KZ scaling~\cite{Zurek1,Zurek2,revqkz1,revqkz2,Zurek4}. Whether such a scaling exists or not in FOPTs is the question addressed herein. 

We have kept $H$ for generality, its presence polarizes both the disordered and ordered phases and one specific ordered phase is energetically favored against the other. Consequently, no KZ topological defects and their related scaling exist at all. However, we will see below that $H$ is a prerequisite for complete scaling of other properties than the defects. We note that previous studies have indicated thermal scaling~\cite{Rao1,zhongpret,zhongssc,Yildiz,Fan,Lee16,Pelissetto16,Kundu} and theories have been developed~\cite{liange,liang}. However, systematical evidence is lacking.

{\it Theory.}---Because the MF transition can only take place beyond the spinodal, we expand the order parameter near the spinodal value $M_s$~\cite{zhongl05,zhong16}, viz.,
$\phi=M_s+\varphi$.
To the leading orders, one finds
\begin{eqnarray}
f_3(\varphi)=\frac{1}{2}\tau\varphi^2+\frac{1}{3}a_3\varphi^3-h\varphi,\qquad\qquad\qquad\label{f3}\\
	\tau=a_2+3a_4M_s^2+5a_6M_s^4,~ \hat{H}_s=a_2M_s+a_4M_s^3+a_6M_s^5,\nonumber\\
	a_3=3a_4M_s+5a_6M_s^3\qquad\qquad\qquad\qquad\label{tauh}
\end{eqnarray}
where $h=H-H_s$ with $H_s=\hat{H}_s$ in MF. The effective reduced temperature $\tau$ and the effective field $h$ become zero exactly at the MF spinodal because they are just the first and second derivatives of $f_6$ with respect to $\phi$. Indeed, for $H=0$ in cooling, $M_s=M_{s}^{\rm MF}=0$ and $a_{2s}^{\rm MF}=0$ at the MF spinodal. 
Therefore, for $f_3$, $\tau=0$ and $h=0$ exactly at the MF spinodal, similar to $a_2=0$ and $H=0$ at a MF critical point (e.g., for $a_4>0$ in $f_6$). This indicates that the universal long-wavelength long-time unstable behavior of FOPTs is in fact controlled by the effective cubic theory ($F$ with $f_6$ replaced by $f_3$ and $\phi$ by $\varphi$) near the spinodal where higher-order terms are irrelevant in the sense of renormalization-group theory~\cite{Mask,Cardyb,Justin,Amit}. 

When fluctuations are taken into account, we have seen in the field-driven FOPTs~\cite{zhong24,zhong24f} and will see below that we can still have a spinodal which is displaced from the MF one by fluctuations, even though the system considered is of short-ranged interactions~\cite{Gunton83,Binder,Binder2}. This validates the effective cubic theory. Standard renormalization-group theory then yields~\cite{zhongl05,zhong16}
\begin{equation}
	M-M_s=b^{-\beta/\nu}g(\tau b^{1/\nu}, hb^{\beta\delta/\nu}, tb^{-z},a_3^*b^{\epsilon/2}),\label{mheq}
\end{equation}
controlled by a nontrivial fixed point $a_3^{*2}=-\epsilon b^{-\epsilon}/6N_d$ below the upper critical dimension $d_c=6$, where $\epsilon=6-d$, $N_d$ is a $d$-dependent constant, $b$ a length scale, $g$ a universal scaling function, and $\beta$, $\delta$, $\nu$, and $z$ are the cubic counterparts of the critical exponents with identical symbols and meaning~\cite{zhongl05,zhong16}. We note that the fixed point is imaginary. However, it has been shown that the renormalization brings the system to an unstable point by removing modes for nucleation so that $a_3$ needs to be analytically continued to imaginary and the system automatically converges to the fixed point~\cite{zhong12,zhong16}. Above $d_c$, $\epsilon<0$ and a Gaussian fixed point takes over. Equation~(\ref{mheq}) is still valid and MF behavior appears with $z=2$, $\nu=1/2$, $\beta=1$, and $\delta=2$~\cite{zhongl05,zhong16,Zenged}. In Eq.~(\ref{mheq}), $\tau$ and $h$ need to include fluctuation contributions for $d<d_c$ similar to the reduced temperature in critical phenomena~\cite{Mask,Cardyb,Justin,Amit}.

We now relate the parameters of the effective cubic theory to those of the original $f_6$. The important point to note is that complete scaling demands all parameters be properly rescaled to achieve full curve collapse~\cite{zhong24}. This necessitates the analysis of their scaling dimensions. There exist two sets of parameters. The first set has simple relationship. Because of Eq.~(\ref{tauh}), $a_4$ and $a_6$ scale as $a_4b^{[a_4]}$ and $a_6b^{[a_6]}$, similar to Eq.~(\ref{mheq}), with the scaling dimension (denoted by square brackets) $[a_4]=[a_3]-[M_s]=\epsilon/2-\beta/\nu$ and $[a_6]=\epsilon/2-3\beta/\nu$, respectively. Also $T$ scales as $T b^{[T]}$ with $[T]=z+2\beta/\nu$~\cite{zhong24}. All can also be directly obtained from a dimension analysis to Eq.~(\ref{lang2d})~\cite{zhong24}. 
Another set of parameters comprises $a_2$ and $H$ which relate to $\tau$ and $h$ through Eq.~(\ref{tauh}).  Note that $[a_4M_s^2]=[a_6M_s^4]=\epsilon/2+\beta/\nu$, $[a_2M_s]=2+\beta/\nu$, and $[a_4M_s^3]=[a_6M_s^5]=\epsilon/2+2\beta/\nu$. Two different situations appear. In MF, valid above $d>d_c$, $d$ is confined in the effective dimension $d_c$~\cite{Zenged} and $\epsilon=0$. The MF exponents then ensure that all terms in each of Eq.~(\ref{tauh}) share identical dimensions. Therefore, $\tau$ and $h$ in Eq.~(\ref{mheq})
can be replaced by $a_2$ and $H$ directly. However, in $d<6$, all quantities containing $M_s$ such as $a_4$ and $a_6$ have dimensions different from $\tau$ and $h$ and are thus singular, because $\nu\neq1/2$ and $\delta\neq2$ even though $\beta=1$ for $d<6$. Similarly, nontrivial fluctuation contributions arising from $a_4$ and $a_6$ are also singular, though those from the cubic theory are not~\cite{zhong24}. Accordingly, the former cannot be ignored while the latter can.

To proceed, we need to specify the KZ condition. This is to cool the system through $a_2=a_{2{\rm i}}-Rt$ with a constant rate $R$ and an initial value $a_{2{\rm i}}$. By proper choice of the time origin so that $\tau=Rt$ (using the same $t$), one finds the dimension of $R$, $r\equiv[R]=z+1/\nu$~\cite{zhong05}. Therefore, setting $b=R^{-1/r}$, we arrive at the finite-time scaling (FTS) forms~\cite{Gong,Gong1,Huang}
\begin{eqnarray}
		M=R^{1/2}g_1(a_2R^{- {1/2}},
		HR^{-1}, TR^{- 3/2},a_4R^{1/2},a_6R^{3/2}),\nonumber\\\label{phigrmf}\\ 
		\begin{split}
			M=R^{\beta/r\nu}g_2\big((a_2+3a_4M_s^2+5a_6M_s^4+\delta a)R^{- {1/r\nu}},\qquad\\
			(H-\hat{H}_s+\delta H)R^{- {\beta\delta/r\nu}},TR^{- [T]/r},\qquad\\
			a_4R^{-[a_4]/r},a_6R^{-[a_6]/r}\big),\qquad\label{phigr}	
		\end{split}
	\end{eqnarray}
for the MF and $d<d_c$, respectively, where $g_i$ is a scaling function, $\delta a$ and $\delta H$ account for all fluctuation contributions excluding those from the cubic theory, and we have replaced $a_3$ with $a_4$ and $a_6$. Note that we do not need to subtract $M_s$ and $H_s$ themselves in Eqs.~(\ref{phigrmf}) and~(\ref{phigr}) since they just contribute an overall displacement of the rescaled curve. Note also that $b^{z}=R^{-z/r}$ is a controllable finite timescale that serves as the temporal analogy of a finite system size in finite-size scaling~\cite{Amit}. In contrast to the KZ scaling that arises from adiabatic approximation, FTS can describe both adiabatic and diabatic regimes~\cite{Huang} and has also been successfully applied to a lot of systems~\cite{Yin,Yin3,Gong1,Liu,Huang,Liupre,Liuprl,Pelissetto,Yinl,Xu,Xue,Feng,Cao,Zhai,Gerster,Li,Mathey,Clark,Keesling,Kang,Yuan,Rossini,Oshiyama,Kuo,Tara,Bacsi,Soto}. In Eq.~(\ref{phigr}), we have also neglected corrections to scaling including deviation from the fixed point~\cite{Wegner,zhong16,Justin,Amit}. 
 
We use curve collapse to verify Eqs.~(\ref{phigrmf}) and~(\ref{phigr}). To this end, all arguments but one of $g_i$ must be fixed. Therefore, for a reference curve run with all parameters denoted with subscript $0$, we choose $a_{2}=a_{2{\rm i}}-Rt$ and
\begin{eqnarray}
\begin{split}
a_4=a_{40}(R_0/R)^{-[a_4]/r},\qquad a_6=a_{60}(R_0/R)^{-[a_6]/r},\qquad\\
T=T_0(R_0/R)^{-[T]/r},\qquad M_s=M_{s0}(R_0/R)^{-\beta/r\nu},\qquad\label{para}\\
\end{split}\\
\begin{split}
	H=\left\{H_0+\delta\hat{H}-a_{2}M_{s0}\left(1-(R_0/R)^{\rho_1}\right )(R_0/R)^{1/r\nu}\right.\qquad\\
		-(a_{40}M_{s0}^3+a_{60}M_{s0}^5)\left(1-(R_0/R)^{\rho_2}\right)\Big\}(R_0/R)^{-\beta\delta/r\nu},\quad~~\label{h}
\end{split}
\end{eqnarray}
with $\delta \hat{H}=\delta H_{0}-\delta H(R_0/R)^{\beta\delta/r\nu}$, $\rho_1=(\beta\delta-2)/r\nu$, and $\rho_2=-[a_4M_s^3]/r+\beta\delta/r\nu$ by adjusting $M_{s0}$ and $\delta\hat{H}$ so that the new curve overlaps the reference one in the rescaled form of $MR^{-\beta/r\nu}$ versus $\tau R^{-1/r\nu}=\{a_2+(3a_{40}M_{s0}^2+5a_{60}M_{s0}^4)(R_0/R)^{-[a_4M_s^2]/r}+\delta a\}R^{-1/r\nu}$ according to Eq.~(\ref{phigr}). The two $\rho$'s reflect the different dimensions of the first and the last two terms in $\hat{H}_s$ and thus must be subtracted to achieve scaling. Similarly, the last two terms in $\tau$ must be deducted along with the fluctuation contributions $\delta a$ again because of their singularity. In MF, Eq.~(\ref{h}) correctly reduces to $H=H_0(R_0/R)^{-\beta\delta/r\nu}$ because $\rho_1=\rho_2=0$ and $\delta H=0$. Similarly, since $[a_4M_s^2]=2=1/\nu$ and $\delta a=0$ in MF, the additional terms to $a_2$ are a constant and $\tau R^{-1/r\nu}$ is equivalent to the direct MF result of $a_2R^{-1/r\nu}$ in Eq.~(\ref{phigrmf}). Note that all the fluctuation contributions and $M_{s0}$ can in principle be theoretically determined using loop expansions~\cite{zhong16,Justin,Amit}, though exact results are hard to obtain.

We have derived the scaling theory in FOPTs using the order parameter $M$. Other properties must also show scaling accordingly. For the KZ scaling, one replaces $M$ by the correlation length $\xi$ and the first $R$ factor on the right by $R^{-1/r}$, a driven length scale~\cite{Huang}, resulting in the scaling form of $\xi$. If there would exist two or more degenerate ordered phases, this form would then give rise to the KZ scaling of topological defects, which is proportional to $\xi^{-d}$. However, one sees from Eq.~(\ref{phigr}) that due to $\hat{H}_s$, one must apply the field $H$, Eq.~(\ref{h}), to achieve complete universal scaling. This field eliminates the KZ topological defects and their related scaling in cooling FOPTs as mentioned. One might keep all the parameters constant except for the cooling $a_2$ in zero field. Then, one cannot keep all the arguments of $g_2$ invariant and thus $\xi$ can no longer be only proportional to $R^{-1/r}$. Consequently, such a KZ scaling can only obtain nonuniversal effective exponent. This is also the reason why previous studies yield scattered exponents. In addition, for fixed parameters and hence fixed $M_s$, $h$ varies with $a_2$ and can be tuned to $h=R't$ with $R'=RM_s$ or $r'\equiv[R']=[R]+\beta/\nu$. This implies that $R$ can be replaced with $R'^{r/r'}$, leading to the field-like thermal class in which all exponents equal the field-driven ones~\cite{liang}. This further complicates the situation.

{\it Method.}---To test the theory, we solve Eq.~(\ref{lang2d}) by direct Euler discretization. Other methods may be more efficient but ought to yield the same results. In the zero-dimensional case [Eq.~(\ref{lang2d}) in the absence of the gradient], the time step is $0.001$ having been checked to bring stable results. The average is over $50000$ samples. The same discretization is also applied to a two-dimensional system. The space step is fixed to 1, while the time step varies from $0.01$ for the small rates to $0.00005$ for the largest rate, again checked to be sufficient. The lattice is $200\times200$ with periodic boundary conditions. Doubling the size has little effect. More than $5000$ samples are employed for average. The initial state is checked to have no effect once it is sufficiently far away from the transition.

\begin{figure}
\centerline{\includegraphics[width=\linewidth]{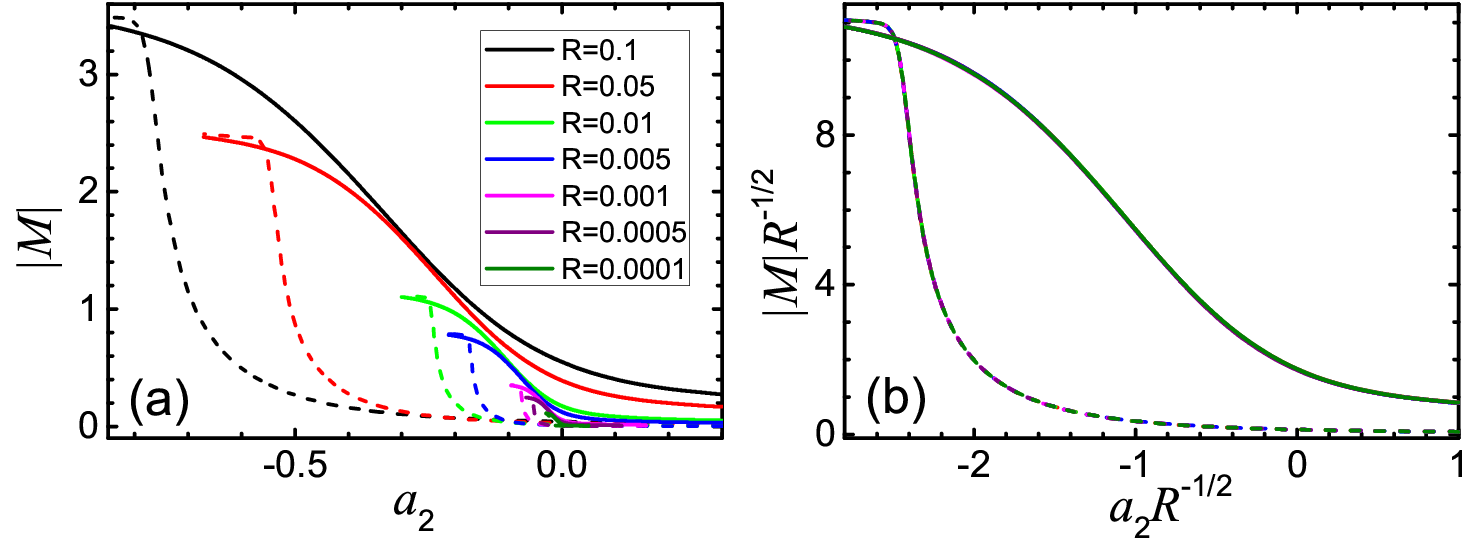}}
\caption{\label{tm0d}(Color online)
(a) $M$ versus $a_2$ for a series of cooling rate $R$ (listed in the legend) from numerical solutions of Eq.~(\ref{lang2d}) in the absence of the gradient term. Dashed and solid lines stand for $T=0$, $H_0=0.01$ and $T_0=0.0025$, $H_0=0$, respectively. $R_0=0.01$, $a_{40}=-1$, and $a_{60}=1$. (b) Rescaling of all curves in (a) according to Eq.~(\ref{phigrmf}). The noiseless case has $|M|=M$ because $H_0>0$ and $M>0$.}
\end{figure}
{\it Results.}---We first consider the zero-dimensional case, which has been shown to control by the MF exponents even in the presence of the noise~\cite{zhong24}. In the noiseless case, a small $H_0$ is needed for the transitions that happen beyond the spinodal near $a_2=0$. This lifts the degeneracy of the ground states and only one ordered phase emerges. With a finite noise, $H_0$ can be zero and the transitions begin before $a_{2s}^{\rm MF}=0$, see Fig.~\ref{tm0d}(a). Both the ordered phases appear and absolute values of $M$ are needed. One sees that all curves in each case in Fig.~\ref{tm0d}(a) perfectly collapse onto a single curve in Fig.~\ref{tm0d}(b) in complete compliance with Eq.~(\ref{phigrmf}) according to the choice of the parameters. This is complete universal scaling. The scaling collapses in Fig.~\ref{tm0d}(b) cover three orders of magnitude of $R$ and have no limitation, firmly confirming the theory. The noisy case may be regarded as perfectly obeying the KZ scenario although no defects are generated since the system has no spatial extension. Note that the exponents used stem from the cubic theory $f_3$ rather than any other continuous transitions. This is not trivial because the transition itself originates from $f_6$ instead of $f_3$. These results indicate the essential role of the spinodal even though it may be considered to be nonexistent for the noisy case~\cite{zhong24,zhong24f}.

Next we study two-dimensional systems in which $\nu=-5/2$, $\delta=-6$~\cite{Cardy85}, $\beta=1$ all exactly, reducing uncertainty in higher dimensions, and $z=1.85$~\cite{zhong24}. Here, an external field $H$, Eq.~(\ref{h}), must be applied due to the singular $\hat{H}_s$, even if $H_0$ is zero. This finite $H$ again eliminate the KZ topological defects and their related scaling. We thus study scaling of the order parameter.

We still adhere to the cooling transition from the disordered phase to the ordered one. As $H$ also varies with $a_2$, its slope determines whether the system transitions from the disordered phase to $M_{\rm eq}$ or to $-M_{\rm eq}$, referred to as the positive phase and negative phase, respectively. Because these two transitions have different $M_{s}$, we focus on the first one with a positive $M_{s}$, though the other one is parity symmetric. This requires $H$ increases from negative to positive as the time evolves and thus $R$ must be larger than $R_0$, different from the zero-dimensional case above. Note that there is a field-driven transition between the two ordered phases owing to the varying field, a transition having its own $M_s$ and must be separated.  

We choose $R_0=0.00008$ for the reference curve. A small $H_0$ is required for the transition to the positive phase. To determine $H$, Eq.~(\ref{h}), $M_{s0}$ is needed, which falls within the turning part of the reference curve. This is affected by $H_0$, which has two effects. One is that the slope of a curve decreases as $H$ increases. The other is that only sufficiently large $H$ for the large rates can have a transition from the polarized disordered phase to the positive phase; otherwise the system directly transitions from the negative phase to the positive phase. Therefore, $H_0$ cannot be too small. Given $H_0$ and $M_{s0}$, for another $R$, there exists a definite $\delta\hat{H}$ such that the new curve is parallel to the reference one owing to the regular slope change. Finally, an equally definite displacement $\delta a$ of the new curve is performed to collapse the two curves onto each other.

\begin{figure}
	\centerline{\includegraphics[width=\linewidth]{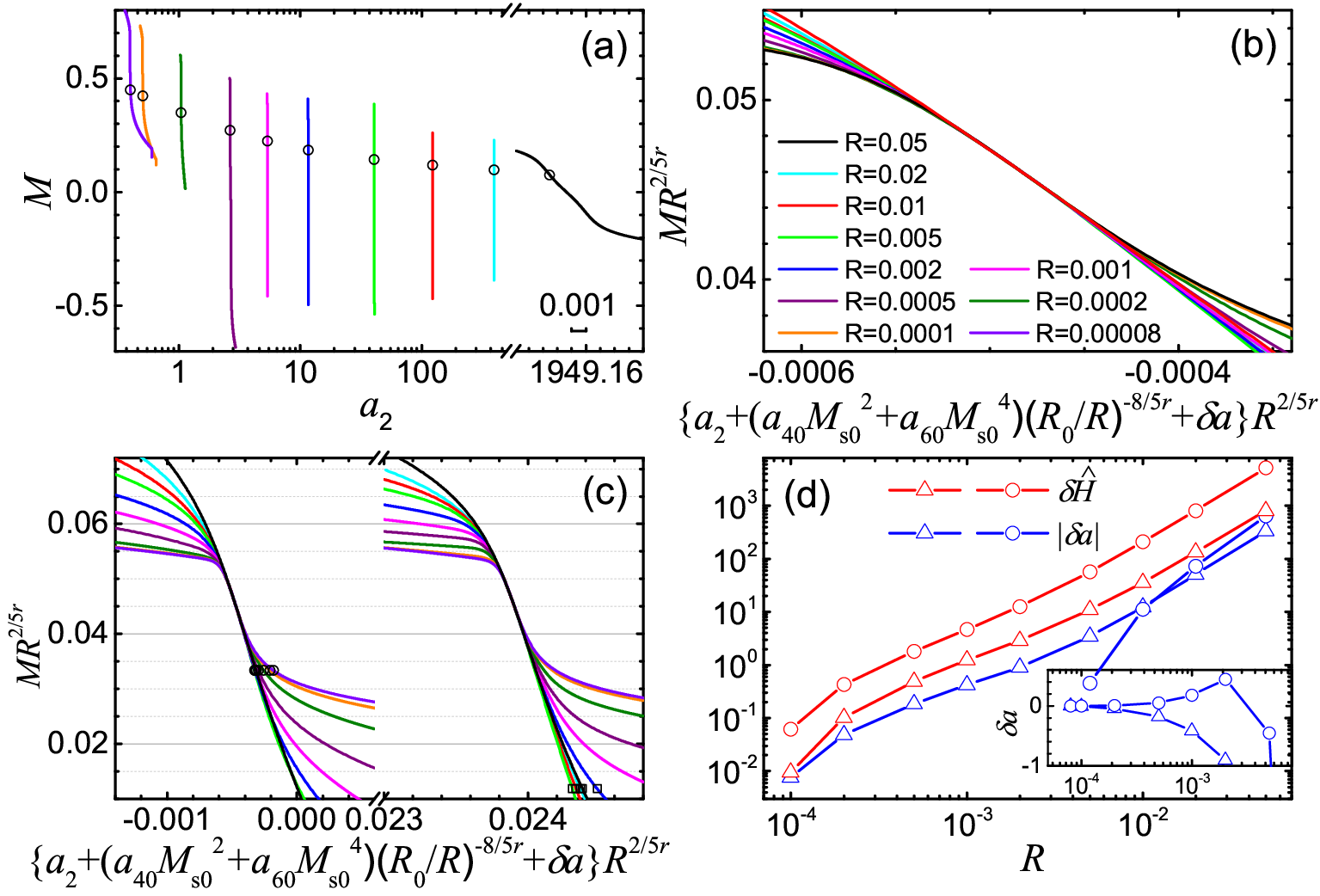}}
	\caption{\label{tm2d}(Color online)
		(a) $M$ versus $a_2$ for a series of $R$ [listed in (b)] from numerical solutions of Eq.~(\ref{lang2d}) in two dimensions. $a_{40}=-1$, $a_{60}=1$, $R_0=0.00008$, $H_0=0.0974$, $T_0=0.01$ for the reference curve (leftmost) and Eqs.~(\ref{para}) and~(\ref{h}) with $M_{s0}=0.45$ for the other curves. The circle on each curve marks its $M_s$. The corresponding $a_2$ (not necessarily $a_{2s}$) is only an interpolation. (b) Rescaling of all curves in (a) according to Eq.~(\ref{phigr}). The curve collapse is good except for crossovers (see the text).   (c) Another perspective of the collapsed curve in (b) (left) and another collapses for $M_0=0.16$ (right with the squares marked $M_{s}R^{-1/r\nu}$). (d) $R$ dependence of $|\delta a|$ (blue) and $\delta{\hat H}$ (red) for $M_0=0.45$ (circles) and $M_{s0}=0.16$ (triangles). The inset shows $\delta a$ near 0. Lines connecting symbols are only a guide to the eye.}
\end{figure}
In Fig.~\ref{tm2d}(a) we draw $M$ versus $a_2$ for a series of $R$ with the corresponding $\delta\hat{H}$ displayed in Fig.~\ref{tm2d}(d) and $M_s$ marked. The $a_2$ range is huge even in a logarithmic scale. For the smallest rates, the transition is directly from the polarized disordered phase to the positive phase, while the remaining rates give rise to a transition first from the negative phase to the polarized disordered phase and then to the positive phase, as can be seen from both the positions of the lower parts of the curves and the enlarged $R=0.05$ curve. This then requires a progressively increasing $\delta\hat{H}$ to balance the rapidly varying $H$, seen in Fig.~\ref{tm2d}(d). After being rescaled according to Eq.~(\ref{phigr}) with the displacement $\delta a$ also depicted in Fig.~\ref{tm2d}(d), all the curves in Fig.~\ref{tm2d}(a) perfectly collapse onto a master curve in Fig.~\ref{tm2d}(b) except for the equilibrium and metastable states and their respective crossovers, which are not expected to be described by the cubic fixed point~\cite{zhong24}. $\delta a$ becomes more negative to cancel the effect of large $H$ for large rates, as Fig.~\ref{tm2d}(d) shows. It does not manifest simple power-law with $R$ since it is the sum of many different power-laws in loop expansions. Neither does $\delta\hat{H}$. The non-monotonic variation of $\delta a$ arises from the nonlinear dependence on $M_{s0}$ of the singular part in $\tau$, which may also be inferred for the smaller $M_{s0}$ used below.

As $M_{s0}$ is not exactly known, we choose another $M_{s0}=0.16$. The collapsed curve is shown in Fig.~\ref{tm2d}(c) (right) and the corresponding $\delta\hat{H}$ and $\delta a$ are also depicted in Fig.~\ref{tm2d}(d) (triangles). The collapsed portion is the same as that of $M_{s0}=0.45$, though the rescaled $M_s$ (marked by squares) is far lower than the other one (circles). The two collapsed curves are parallel because the reference curve is identical and the displacement off $a_2$ is a constant. However, different $M_{s0}$ values change the slope of $H$ versus $a_2$. Consequently, although a change of $M_{s0}$ can be simply accounted by a related change of $\delta a$, it must be balanced by a varying $\delta\hat{H}$. A constant $\delta\hat{H}$ then leads to a different set of curves, as is visible in Fig.~\ref{tm2d}(c).  Nevertheless, owing to $\delta a$ and $\delta\hat{H}$, the same collapse for the two different $M_{s0}$ values indicates that they do not affect but confirm the scaling. Finite-size effects may also be so accounted for.

The collapses in Fig.~\ref{tm2d} have already covered more than two orders of magnitude. In fact, the range of $R$ is not limited. The only issue is time. For larger $R$ values, $H$ changes so steeply with $a_2$ that very small time steps are needed to solve Eq.~(\ref{lang2d}), while smaller $R$ values need longer computation time. We note that the parameters chosen for Eq.~(\ref{lang2d}) are generic. It involves four objects, $t$, $\bf{x}$, $H$, and $\phi$. They can then be properly scaled to fix four parameters, having been chosen to be $a_{40}$, $a_{60}$, $H$ itself, and the coefficient of the gradient. As $a_2$ decreases with cooling, only $T$ is adjustable. If $T$ is too small, fluctuations are negligible and the MF results recover~\cite{zhong24}; while if it is too large, a new transition occurs~\cite{zhongpre}. Within this range, any $T$ should give rise to the same result as the one chosen is not special. Therefore, the good curve collapse within such a large range through tuning only just three parameters $M_{s0}$, $\delta a$ and $\delta\hat{H}$ (in fact only just one since the other two are fully determined by the first one) indicates that there must exist some underlying scale invariance even if there were no theory. 

{\it Conclusion.}---We have demonstrated that for a cooling first-order phase transition, the order parameter exhibits complete universal scaling except for two crossovers to equilibrium and metastable states. However, as the system completely transforms from a polarized disordered phase to an ordered phase owing to an intrinsic symmetry-broken field responsible for the scaling, no topological defects of the KZ mechanism are generated and thus any KZ scaling for the topological defects can only be a rough approximation.

\begin{acknowledgments}
I thanked Dr. Shuai Yin for bringing Ref.~\cite{Suzuki} to my attention. This work was supported by National Natural Science Foundation of China (Grant No. 12175316).
\end{acknowledgments}

\end{document}